4 May 2021

# Deep learning-based tumor segmentation on digital images of histopathology slides for microdosimetry applications

Luca L. Weishaupt[1], Jose Torres[2], Sophie Camilleri-Broët[2], Roni F. Rayes[3], Jonathan D. Spicer[3], Sabrina Côté Maldonado[1], Shirin A. Enger[1,4]

[1] Medical Physics Unit, Department of Oncology, Faculty of Medicine, McGill University, Montréal, Québec, Canada

[2] Department of Pathology, Faculty of Medicine, McGill University, Montréal, Québec, Canada

[3] Cancer Research Program and the LD MacLean Surgical Research Laboratories, Department of Surgery, Division of Upper GI and Thoracic Surgery, Research Institute of the McGill University Health Center, Montréal, Québec, Canada

[4] Research Institute of the McGill University Health Center, Montréal, Québec, Canada

**Corresponding Author:**
Luca Weishaupt
E-mail address: luca.weishaupt@mail.mcgill.ca

## Abstract

**Purpose:** The goal of this study was (i) to use artificial intelligence to automate the traditionally labor-intensive process of manual segmentation of tumor regions in pathology slides performed by a pathologist and (ii) to validate the use of a well-known and readily available deep learning architecture. Automation will reduce the human error involved in the manual process, increase efficiency, and result in more accurate and reproducible segmentation. This advancement will alleviate the bottleneck in the workflow in clinical and research applications due to a lack of pathologist time. Our application is patient-specific microdosimetry and radiobiological modeling, which builds on the contoured pathology slides.

**Methods:** A deep neural network named U-Net was used to segment tumor regions in pathology core biopsies of lung tissue with adenocarcinoma stained using hematoxylin and eosin. A pathologist manually contoured the tumor regions in 56 images with binary masks for training. To overcome memory limitations overlapping and non-overlapping patch extraction with various patch sizes and image downsampling were investigated individually. Data augmentation was used to reduce overfitting and artificially create more data for training.

**Results:** Using this deep learning approach, the U-Net achieved accuracy of 0.91±0.06, specificity of 0.90±0.08, sensitivity of 0.92±0.07, and precision of 0.8±0.1. The F1/DICE score was 0.85±0.07, with a segmentation time of 3.24±0.03 seconds per image, thus achieving a 370±3 times increased efficiency over manual segmentation, which took 20 minutes per image on average. In some cases, the neural network correctly delineated the tumor's stroma from its epithelial component in tumor regions that were classified as tumor by the pathologist.

**Conclusion:** The readily available U-Net architecture can segment images with a level of efficiency and accuracy that makes it suitable for tumor segmentation of histopathological images in fields such as radiotherapy dosimetry, specifically in the subfields of microdosimetry.

## I.  Introduction

Tissue response to ionizing radiation has been shown to vary between different tissue and tumor types as well as between different radiation qualities, which are defined by the type and energy of the radiation.

The field of microdosimetry investigates potential sources explaining the differences in tissue responses following exposure to ionizing radiation with different radiation qualities. It involves studying fluctuations of local energy absorptions at the subcellular level from single tracks of ionizing radiation. An underlying assumption of microdosimetry is that energy deposition in sensitive sites such as the cell nucleus and radiation effects are correlated.[4,5] The probability of obtaining a certain amount of energy absorbed in a target volume depends on the size of the target volume as well as two stochastic phenomena: i) the number of single tracks passing through the volume and ii) the amount of energy that the tracks impart in the volume. Due to the presence of this microdosimetric spread, individual target volumes such as cell nuclei do not necessarily receive the same amount of absorbed energy, when ionizing radiation traverses through the tissue.[6,7]

The size distributions of cells and nuclei depend on various factors: the tissue type, cell cycle, and the malignancy, all of which also vary between patients. Tumor nuclei generally have more variability in size and shape compared to healthy nuclei.[8] Tumor cell nuclei subject to histopathological preparations confirm the variability in size, shape, and chromatin patterns from that of healthy tissue and each other.[9] Commonly, the nucleus of a tumor cell appears more substantial and darker than that of a corresponding normal cell due to an excess of DNA content within its volume.

Currently, in microdosimetry calculations performed with the Monte Carlo (MC) method, properties such as energy absorption have been studied using a target geometry made up of single-sized volume mimicking individual cells.[10–16] To represent nucleus sized targets, some authors use smaller volumes, which reside within the cell body.[11,13] While using uniformly sized cells as a target geometry may provide useful information in many dosimetry applications, any effects that a nuclei population with diverse volume sizes may have on the studied quantities are usually ignored due to lack of available data. The approximation of uniformed sized nuclei and cells overlooks the microdosimetric spread in energy deposition due to the stochastic nature of ionizing radiation interaction with matter.[17] As the microdosimetric spread directly influences the dose-response and affects the treatment outcome, there exists a need for the consideration of nuclei/cell size distributions in dosimetry investigation and radiobiological modeling of tissue response.[18,19]

To study the effect of varying size distributions of cells and nuclei, we have previously developed a comprehensive software package to build morphologically realistic three-dimensional cellular geometry models.[20] These models produced by the software are used as input data to MC based dose calculations for patient-specific dosimetry, including microdosimetry and theranostics.[3] The developed software creates these three-dimensional models by deriving cell and cell nuclei size distributions from digital images of 2D histopathological slides, which were contoured manually by a pathologist.

However, the manual contouring done by a pathologist is tedious, expensive, and time-consuming, creating a bottleneck not just in the workflow of patient-specific microdosimetry investigations and radiotherapy but also other scientific fields requiring contoured tissue microarrays such as immunotherapy.[21–25] It takes a pathologist 20 to 30 minutes to contour a single image, which involves inspecting large patches of healthy cells to identify malignant regions. Furthermore, manual contours are prone to human error as there are inter- and intra-observer discrepancies.[21,22,26,27] These limitations demonstrate the need for an efficient, accurate, and reproducible method for contouring pathology slides.

A solution to this challenge would be to use an automated process that could perform the segmentation with a high level of accuracy. The semantic segmentation of histopathological images is a prevalent computer vision problem, and various methods have been proposed to perform it. An approach that has shown great promise in recent years is the use of deep convolutional neural networks,[28] which can learn high-level feature information needed for pathology image segmentation. Deep learning algorithms of many forms have been used for several types of histopathological segmentation, including the segmentation of tumor regions.[28]

The analysis of labeled histopathological images presents various challenges including their large dimensions, the inter- and intra-observer variability in the labels created by pathologists, and the scarcity of accurately labeled data.[29] Methods have been developed to combat these challenges and various deep learning methods exist that can perform tumor segmentation on histopathological images with a high level of accuracy. [30–32]

Ronneberger *et al*. (2015) developed a deep learning architecture that can perform pixel level semantic segmentation of medical images.[33] However, traditionally in histopathology there is a lack of high quality training data. Usually, manual labels are inconsistent and not created at pixel level precision. To address these challenges Chen *et al*. (2019) and Linmans *et al*. (2018) investigated data augmentation in the form of random noise and permutations to the image data.[31,32] However, another challenge presented by the histopathological images is their large size, which makes their analysis computationally expensive. To overcome this Wang *et al*. (2019) investigated the use of image-level labels with a weakly supervised deep learning model to perform whole slide image classification.[30]

The goal of this study was not to outperform these levels of accuracy but to validate the viability of using the popular and relatively simple U-Net architecture[33] to segment tumor regions in histopathological images for microdosimetric investigations. Although training the algorithm may be computationally expensive, the computational requirements for using the trained model should be low enough to perform the segmentation on most computers and without access to a supercomputer.

To minimize manual labor and human error in the segmentation of the tumor regions in images of biopsy slides we validate a deep convolutional neural network model using a U-Net architecture. By automating the traditionally labor-intensive manual segmentation, the dependency on a pathologist in contouring is removed, therefore facilitating the investigation of segmented tumor cells and nuclei for patient-specific dosimetry studies, including microdosimetry and theranostics.

## II.    Methods

### II.I Problem formulation

The segmentation of tumor regions in digital histopathological images was described as a supervised learning problem. Given a set of paired data $\{\mathbf{x}, \mathbf{y}\}$, the objective was to learn a function mapping $\mathbf{f}$ which takes an input $\mathbf{x}_i$ and produces an output prediction $\mathbf{y'}_i$, i.e. $\mathbf{y'}_i = f(\mathbf{x}_i)$, that is an accurate representation of the corresponding $\mathbf{y}_i$ from the set. This should hold for any pair of $x_i$ and $y_i$ in the set.

In this case, $\mathbf{x}$ is a set of two-dimensional histopathological images, and $\mathbf{y}$ represents the set of contour mask created by a pathologist, which describe the location of tumor regions in the corresponding histopathological images. The proposed function $\mathbf{f}$ is a deep convolutional neural network that takes a histopathological image as its input and creates a contour mask as its output. The output is the resulting prediction $\mathbf{y'}$. The neural network achieves this through repeated convolutions using weights and biases optimized on a training subset of $\{\mathbf{x}, \mathbf{y}\}$.

### II.II Data Set

This retrospective study received approval from our institutional review board at McGill University Health Center. We received informed consent from patients for their participation and for the use of their images in the study. All methods were carried out in accordance with relevant guidelines and regulations.

Slides from 56 pathology formalin-fixed and paraffin-embedded tissue microarrays cores of lung adenocarcinoma stained by using hematoxylin and eosin were scanned by using an Aperio slide scanner at 40x (Leica Biosystems, Germany). The images had 3750x3750 pixels and were saved in TIFF format. A pathologist (JT) manually contoured the images with a binary mask, distinguishing between tumor and non-tumor regions with an isotropic resolution of 248 nm per pixel on a pixel by pixel basis. The segmentation map was created in XML format and converted to .mat files. The images' RGB channels were normalized by dividing their intensity values by 255.

### II.II.I Data Augmentation

Data augmentation was used to prevent overfitting, improve the robustness of the algorithm, and to artificially increase the amount of training data.[34] Images were randomly rotated over a range of 360 degrees. The height and width of the images were randomly increased or decreased by 20%. Images were randomly slanted with a shear range of 15 degrees. Furthermore, images were zoomed into or zoomed out by 20%, with reflected boundaries in case of zooming out. There was also a 50% chance that images were flipped horizontally and/or vertically. These augmentations were performed after every epoch, meaning after every pass through all the training data.

4 May 2021

## II.II.II Patch Extraction and Downsampling

In an ideal case, the entire image in its native resolution should serve as the input to the neural network, as this would present the most feature and context information to the algorithm. However, there exists a computational limit to how large the input images to a neural network can be. If the size of the image is too large, it becomes a challenge to train the model due to the amount of memory required. While it is theoretically possible to increase the amount of memory allocated to training, this would make the reproducibility of the results difficult.

Hence, since the 120 Gigabytes (GB) of memory available on the CPU supercomputer we had accessed to for this study, were not sufficient to train the proposed neural network using 3750x3750 pixel images, the image size had to be reduced.

Reducing the size of the images could be achieved by either resampling the images to a lower resolution, also known as downsampling, or by extracting patches from the images.

The drawbacks of downsampling are i) image information is permanently lost and will not be used for training and ii) the masks are created in the downsampled resolution are therefore less precise. Extracting patches comes with the caveat of sacrificing contextual information as the patches only contain information about a fraction of the image. This means the algorithm will only "see" one part of the image at a time instead of the whole image.

## II.III Network Architecture

A U-Net architecture was employed for this study.[33] A depiction of the architecture is presented in Figure 1. The model was constructed using Keras with a Tensorflow backend. It consists of two main parts, a contracting path, and an expanding path. There are four contracting steps and four expanding steps. The contracting path extracts high-level semantic information about the tumor regions in the images, while the expanding path restores the spatial information that is lost in the contracting path.

## II.III.I Contracting Path

On the contracting path of the U-Net, the original image patches are fed into the network as three normalized color channels. This input layer is then fed through a convolutional block. A convolutional block consists of two iterations of convolution by a 3x3 kernel followed by a batch normalization layer and a rectified linear unit (ReLU) activation layer. ReLU is used as opposed to similar activation functions such as sigmoid functions because it is more effective for training complex models with large data sets. The output is then fed through a 2x2 max pooling layer with stride 2 to reduce the dimension of the input while preserving relevant features. A dropout layer connects the max-pooling layer to the next convolutional block and drops neurons with a probability of 50% to reduce overfitting. This sequence of a convolutional block followed by a max-pooling layer and a dropout layer is performed four times and makes up the contracting path.

## II.III.II Expanding Path

The final dropout layer from the contracting path feeds into a convolutional block. The output of the convolutional block is then up-convoluted or deconvoluted using the inverse of a convolution with a 3x3 kernel in order to restore the image dimension. The output of the up-convolution layer is fed through a dropout layer and concatenated with a copy of the output of the convolutional block of the layer at the equivalent depth in the contracting path. This is done to preserve information about feature location. The concatenated output feeds into a convolutional block that is then followed by the next up-convolution and concatenation. This sequence of up-convoluting, concatenating, and feeding through a convolutional block is performed four times to create an output with feature maps of the same dimension as the input image. The expanding path ends in a convolution by a 1x1 kernel and a sigmoid function to create a normalized prediction map with the same dimensions as the image patch.

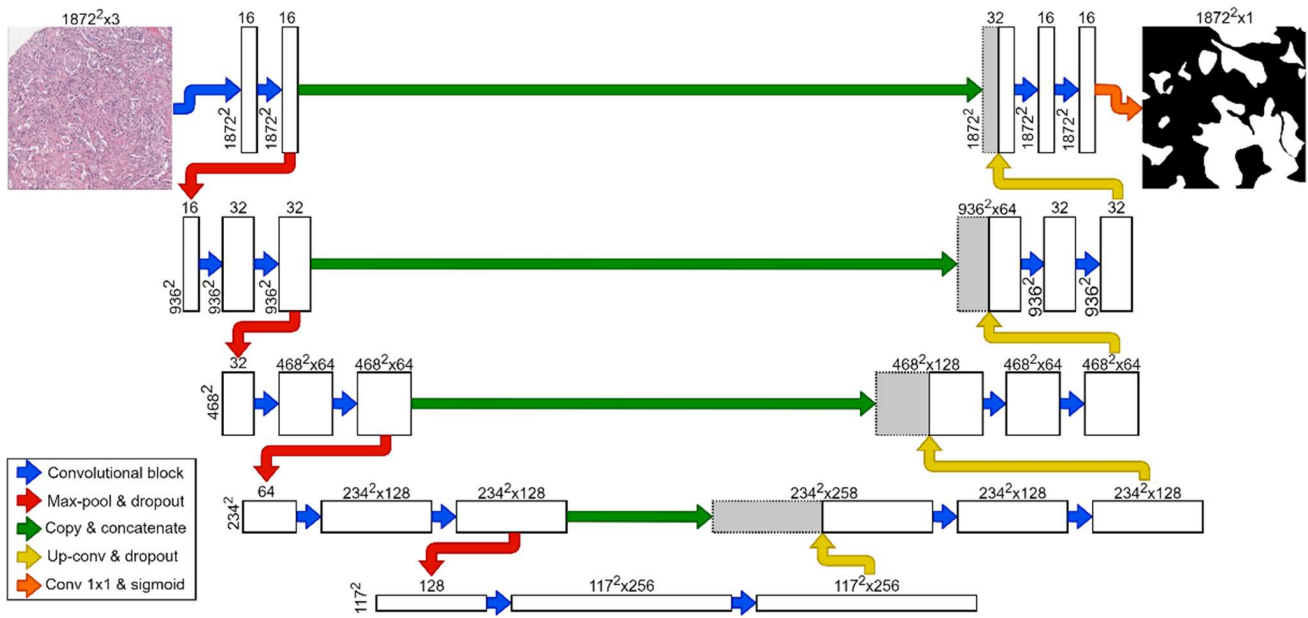

Figure 1: The U-Net architecture employed for image segmentation of 1872x1872 pixel patches. The image patch with its normalized RGB channels undergoes four convolutions, max-pooling, and dropout iterations for feature extraction. The resulting feature maps are then up-convoluted and concatenated with copies of the feature maps from the contracting path to retain location information four times. A convolutional block consists of a convolution with a 3x3 kernel, batch normalization, and a ReLU activation function. In the final step, the feature maps are convoluted with a 1x1 filter and a sigmoid activation function to create a normalized mask for the input image patch.

## II.IV Evaluation Criterion

### II.IV.I 8-fold Cross-Validation

K-fold cross-validation is a statistical method used to determine the accuracy of machine learning models. As the available data is split into a training portion and a testing portion, not all the data can be used for training at the same time. Furthermore, there might be selection bias if the testing data is "simpler" or "more complex" than the rest of the data. It is difficult to determine what is "simple" or "complex" for a deep learning algorithm and thus challenging to select a representative testing batch.

K-fold cross-validation is a method to eliminate any bias and to create a more inclusive representation of a machine learning algorithm's performance. All the available data is split into k folds. Each fold consists of an equal number of images and contours. Then, iteratively, each fold is used as the testing set while the rest of the folds are used as the training set. Hence k separate models are trained and tested. Using the k-fold cross-validation, each image is used for testing once.

In this study, satisfied 8-fold cross-validation was used. Satisfied k-fold validation attempts to create homogenous folds. In this case, the folds were created such that the summed area of all the tumor regions in each fold was roughly the same.

The average reconstruction scores from the eight testing sets were calculated and were reported in the results. Furthermore, the standard deviations were calculated for each of the averages of the reconstruction scores and were reported as statistical uncertainties.

### II.IV.II Loss Function

A soft dice function was used to train the neural network. We define the soft dice function between two masks A and B as

$$SoftDice(A, B) = \frac{2 \sum_{mask} a_i b_i}{\sum_{mask} a_i^2 + \sum_{mask} b_i^2},$$

where $\sum_{mask} a_i$ represents the sum of all pixels in the mask A (similarly for B).

The soft dice loss function, which was used for training in this study, is then simply defined as

$$loss(A, B) = 1 - SoftDice(A, B),$$

where A is the ground truth mask, and B is the mask produced by the neural network. This loss function is convenient for the segmentation of tumor regions because the tumor regions in the images tend to be smaller than the non-tumor regions. The soft dice score is, therefore, a more accurate representation of the algorithm's performance as other scoring metrics such as accuracy, which would award relatively high scores for reconstructed masks with little to no segmented tumors. The soft dice only awards high scores for reconstructions with accurate and confident segmented tumor regions.

## II.V Model Training and Testing

This study compares both downsampling and patch extraction to reduce the image size and memory requirements. To investigate the method of downsampling, all images were downsampled by a factor of 2 using third order spline interpolation, resulting in an image dimension of 1875x1875 pixels. We define this as case 7.

For the patch extraction method, patches of dimensions 1872x1872 pixels, 928x928 pixels, and 464x464 pixels were extracted. The numbers of patches per image were 4, 16, and 64, respectively. The patch dimensions had to be multiples of 16 due to the architecture of the neural network. 1872x1872 pixels was the maximum patch size that could be used for training using 120 GB of memory. The remaining patch dimensions were chosen hierarchically, reducing the patch size by a factor of circa 4 for each dimension.

### II.V.I Training

The model was trained using the hand-contoured masks as ground truth maps and was scored using binary cross entropy as its loss function. 8-fold satisfied cross-validation was performed on the 56 pairs of images and masks made by the pathologist to eliminate selection bias. While training, 20% of the training patches were used exclusively for validation. A batch size of eight was used, which means eight images or image patches were passed through to the model at once. The initial learning rate was 0.001. The learning rate was reduced by a factor of 10 after every interval of 3 epochs without improvements to the validation loss until a minimum learning rate of 0.00001 was reached. The best weights based on validation loss were saved and used for testing. The training was performed on a 120 GB CPU cluster for 24 hours on Niagara supercomputer at the SciNet HPC Consortium.[35,36]

The deep learning algorithm was trained with seven different cases. Six cases were trained to investigate patch extraction (cases 1-6), and one case was trained to investigate the downsampling (case 7). Two cases were trained for each patch dimension, one with overlapping patches and one without overlapping patches. Cases 1 and 2, cases 3 and 4, and cases 5 and 6 correspond to patch dimensions of 1872x1872 pixels, 928x928 pixels, and 464x464 pixels, respectively. The deep learning model architecture and all parameters used except the input dimension and resolution were the same for all test cases.

### II.V.II Testing

During testing, to reduce the effect of the lost contextual information due to patch extraction and to create more continuous contours, overlapping patches were extracted. The patches were extracted with a stride length of 939, 471, and 235 pixels for the 1872-, 928-, and 464-pixel patches, respectively. In every row and every column, a stride of the aforementioned pixel dimension was taken before extracting the next patch, as illustrated in Figure 2. Hence, 9, 49, and 225 patches were extracted from each image, respectively. Overlapping patches were used in the test cases with even indices.

For the images with non-overlapping patches, 4, 16, and 64 patches were extracted from each image for patch sizes of 1872, 928, and 464 pixels, respectively. For each pixel that had more than one prediction due to overlap, the average value from all predictions was taken. Non-overlapping patches were also tested for each patch size for comparison. The patches extracted from the downsampled images were non-overlapping as they had the dimensions as the images. Only non-overlapping patches were used for training as overlapping patches would have created a bias towards the overlapping regions because these regions would have appeared multiple times in the training set. The test cases with odd indices used non-overlapping patches.

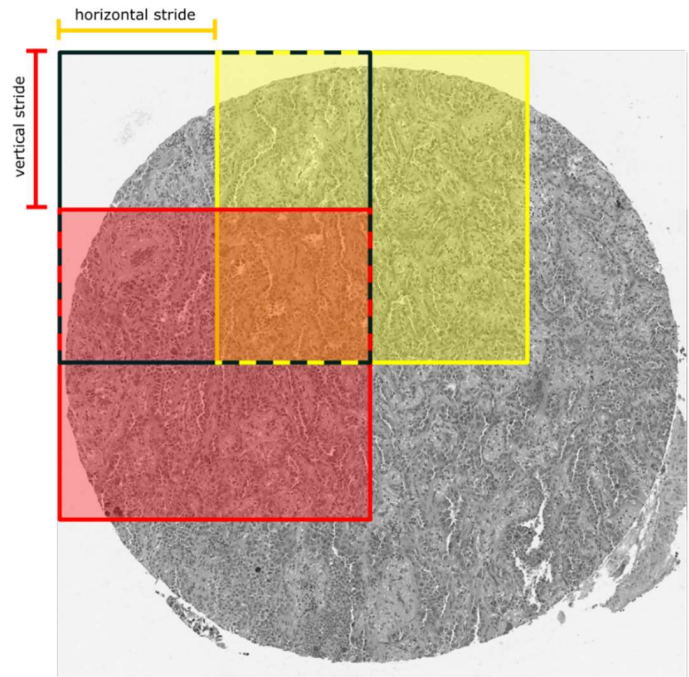

Figure 2: Overlapping patch extraction with 1872-pixel patches. The image was converted to black and white for this illustration. Patch extraction begins at the top left corner. The black square indicates the first patch. A horizontal stride is then taken before the next patch is extracted. The horizontal stride and the subsequent patch are illustrated in yellow. After all patches in one row have been extracted, a vertical stride is taken. The vertical stride and the first patch in the second row are illustrated in red. This process is repeated until the entire image area has been extracted. Nine patches were extracted for this image. Dashed lines are used to indicate overlapping patch edges.

### II.VI Reconstruction Scores

To score the model, the generated prediction maps were binarized using a 0.5 confidence threshold and were then compared to the contours made by the pathologist. The threshold was chosen based on common practice observed in literature and to facilitate comparison with the values from the literature. After applying the threshold, the accuracy, specificity, sensitivity, precision, and F1/DICE score were measured by comparing the U-Net's contours with the contours made by the pathologist. The F1 score and DICE score are the same after a 0.5 threshold is applied. These metrics are defined in Table 1.

| Accuracy | Specificity | Sensitivity | Precision |
|---|---|---|---|
| $\dfrac{TP + TN}{TP + TN + FP + FN}$ | $\dfrac{TN}{TN + FP}$ | $\dfrac{TP}{TP + FN}$ | $\dfrac{TP}{TP + FP}$ |

| F1-score/DICE | Legend |
|---|---|
| $\dfrac{2}{sensitivity^{-1} + precision^{-1}}$ | **TP**: True Positive **TN**: True Negative |
| | **FP**: False Positive **FN**: False Negative |

Table 1: Definition of the accuracy, specificity, sensitivity, precision, and F1/DICE metrics needed for comparison of the contours generated by the U-Net with the contours drawn by the pathologist.

True positives are pixels that were correctly identified as being part of a tumor region. True negatives are pixels that were correctly identified as not being part of a tumor region. False positives are pixels that the reconstruction labeled as being in a tumor region but were labeled as non-tumor by the pathologist. False negatives are pixels that were classified as being part of a tumor region by the pathologist but not by the U-Net.

The accuracy of a reconstruction describes the proportion of pixels that were correctly classified. The specificity measures the fraction of pixels containing no tumor that were correctly classified as such. The sensitivity describes how many of the pixels that were manually classified as being part of a tumor region were correctly classified as such in the reconstruction. The precision is a measurement of the percentage of the correct guesses out of all the pixels that were classified as being part of a tumor region. The F1-sore/DICE coefficient is a measurement of the overlap of the regions that were manually contoured as tumor and the regions that were classified as tumor in the reconstruction.

To attain a threshold-independent score of the algorithms' reconstructions, the area under the receiver operating characteristic curve (ROC) and the area under the precision-recall curve were also measured. These scores give a more holistic representation of the quality of the reconstructed contours. The contours should have a high level of confidence when identifying tumor regions to receive a good score.

## III.    Results

### III.I. Segmentation Time

All segmentation times are presented in Table 2. The average time for the pathologist to contour a core image was 20 minutes. The slowest test case (case 6), had an average segmentation time of $13.1 \pm 0.1$ seconds per image, thus achieving a $91.6 \pm 0.7$ times increased efficiency. The fastest test case (case 7), which was trained on the downsampled data, had an average segmentation time of $3.24 \pm 0.03$ seconds per image, thus achieving a $370 \pm 3$ times increase in efficiency over manual segmentation.

### III.II. Reconstruction scores

Table 2 presents the scores of each test case. The scores for all test cases are within statistical uncertainty of each other. There is a slight increase in the uncertainty for smaller patch sizes.

### III.III. Qualitative Analysis

The prediction maps generated by the neural networks appear visually similar to the hand-contoured region, as shown in Figures 3 and 4. A noticeable difference between the contours is the holes that are present in the large tumor region in the images segmented by U-Net. These holes are not present in the manual contours. As opposed to the mask made by the pathologist, the prediction map generated by the U-Net is non-binary with intensity values indicating the predicted probability of being a tumor region.

The smaller the patches extracted from the images, the less coherent and smooth the reconstructed segmentations appeared. With smaller patches, there is also an increased amount of noise, and the edges of tumor regions are more blurred as presented in Figure 4.

The edges of the extracted patches are visible. In the cases where no overlap was used (cases 1, 3, and 5), the edges are more noticeable and appear to divide tumor regions more than in the cases where overlapping patches were used (cases 2, 4, and 6).

The segmentations that were done on the downsampled images (case 7) have slightly less confidence in identifying the background and healthy regions (Figure 4.7).

### III.IV. Memory Usage

The 464x464 pixel patch test cases (cases 5 and 6) required less than 4 GB of RAM, while the 928x928 pixel patch test cases (cases 3 and 4) required less than 6 GB of RAM. The 1872x1872 pixel patch test case and the test case using downsampled images (case 7) both required less than 8 GB of RAM.

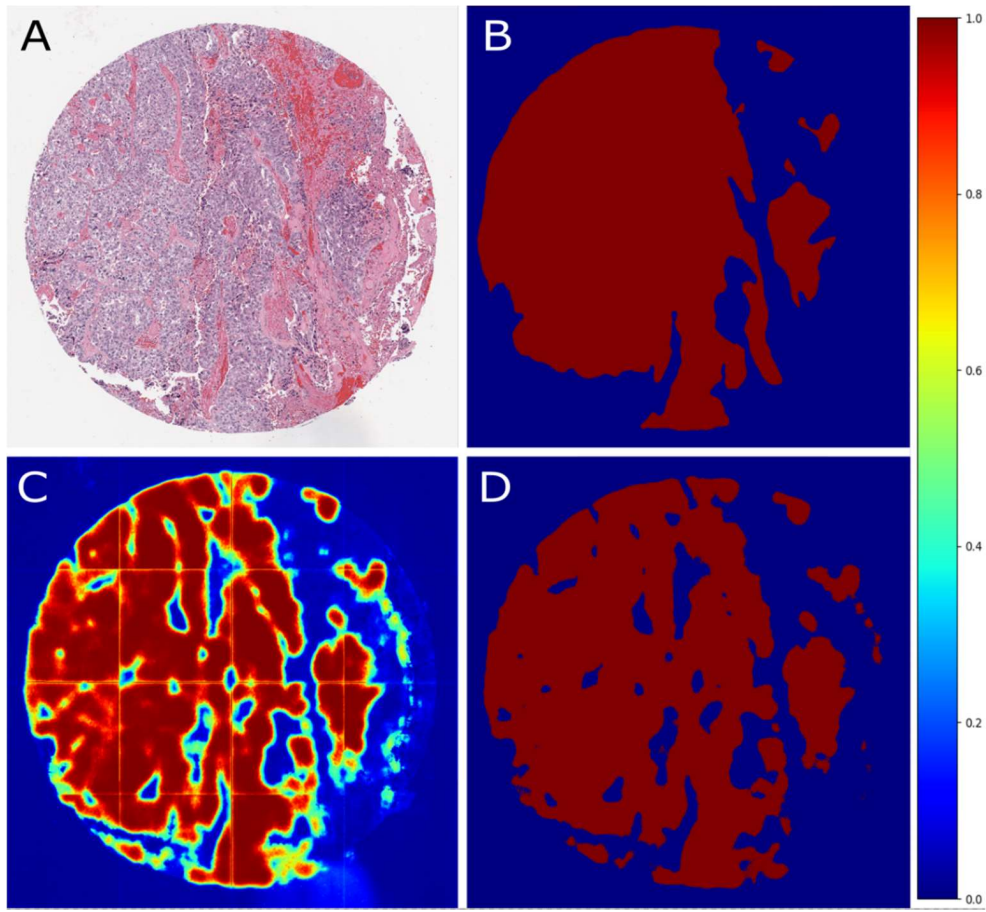

Figure 3: Representative example of a pathology core. A – The original image. B – The contour made by the pathologist. Tumor regions are contoured in red. C – Prediction map generated by the U-Net algorithm with 1872x1872 pixel patches and 939-pixel strides (case 2). D – The prediction map after applying a 50% confidence threshold. High values indicate a high probability of tumor.

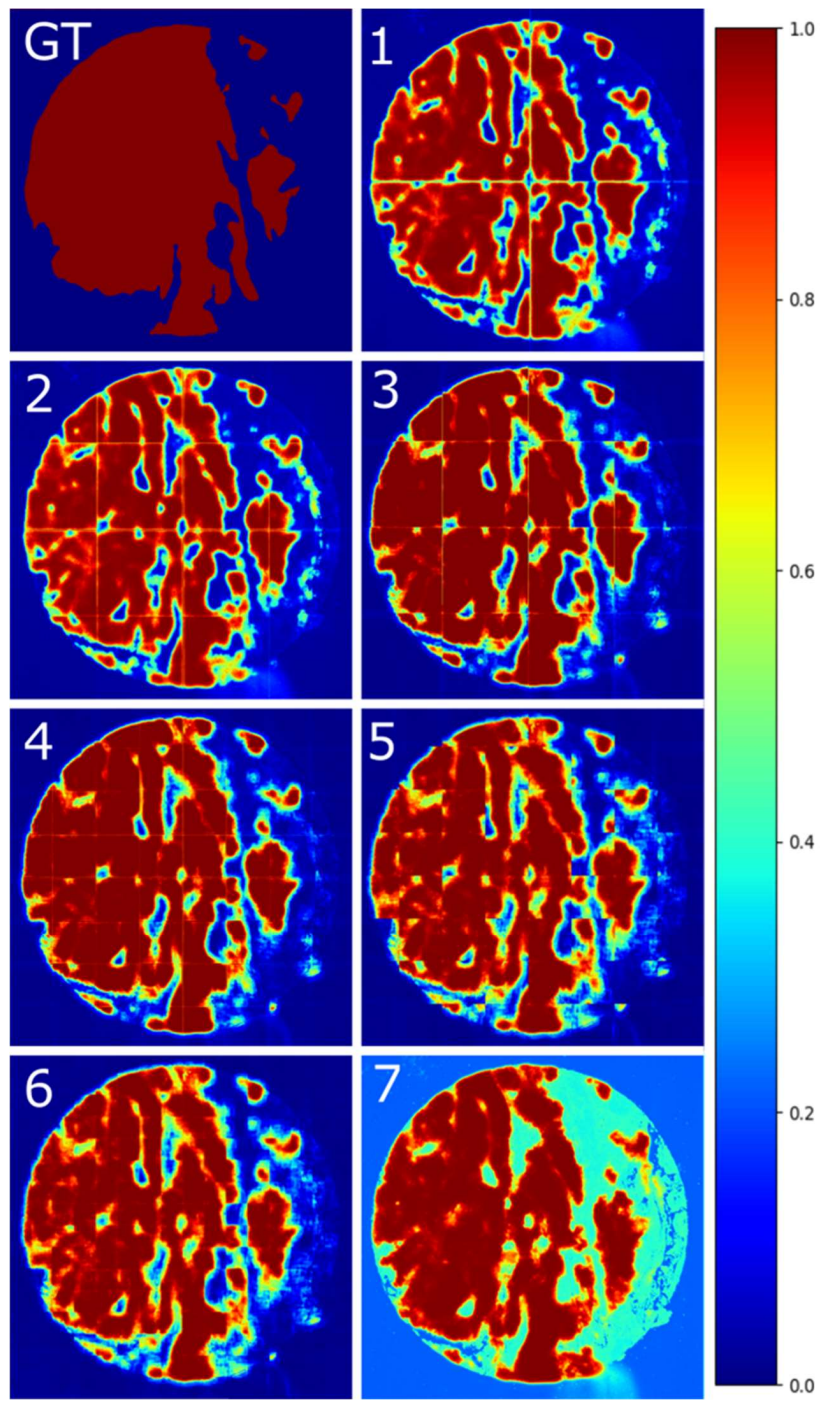

Figure 4: Segmented tumor regions from each of the 7 different test cases. GT is the ground truth contour created by the pathologist. The rest of the indices indicated in each pane correspond with a case index from Table 2.

| Method | Patch size (pixels) | Case Index | Strides (pixels) | Time per image (s) | AUC ROC | AUC Precision-Recall | Accuracy | Specificity | Sensitivity | Precision | F1 score / DICE coefficient |
|---|---|---|---|---|---|---|---|---|---|---|---|
| | | | | | | | *After applying a threshold of 0.5* | | | | |
| Patch Extraction | 1872x1872 | 1 | None | 7.20 ± 0.05 | 0.90 ± 0.05 | 0.86 ± 0.04 | 0.91 ± 0.04 | 0.92 ± 0.06 | 0.9 ± 0.1 | 0.81 ± 0.09 | 0.84 ± 0.05 |
| | | 2 | 939 | 7.20 ± 0.06 | 0.90 ± 0.05 | 0.87 ± 0.04 | 0.91 ± 0.04 | 0.92 ± 0.06 | 0.9 ± 0.1 | 0.81 ± 0.09 | 0.84 ± 0.05 |
| | 928x928 | 3 | None | 5.38 ± 0.06 | 0.89 ± 0.06 | 0.87 ± 0.06 | 0.91 ± 0.05 | 0.93 ± 0.06 | 0.9 ± 0.1 | 0.85 ± 0.08 | 0.84 ± 0.07 |
| | | 4 | 471 | 10.45 ± 0.07 | 0.89 ± 0.06 | 0.87 ± 0.06 | 0.91 ± 0.05 | 0.93 ± 0.06 | 0.9 ± 0.1 | 0.85 ± 0.08 | 0.84 ± 0.07 |
| | 464x464 | 5 | None | 4.78 ± 0.06 | 0.89 ± 0.06 | 0.87 ± 0.06 | 0.91 ± 0.05 | 0.94 ± 0.07 | 0.8 ± 0.1 | 0.9 ± 0.1 | 0.85 ± 0.08 |
| | | 6 | 235 | 13.1 ± 0.1 | 0.89 ± 0.06 | 0.88 ± 0.06 | 0.91 ± 0.05 | 0.94 ± 0.07 | 0.9 ± 0.1 | 0.9 ± 0.1 | 0.85 ± 0.08 |
| Down-sampling | 1872x1872 | 7 | None | 3.24 ± 0.03 | 0.91 ± 0.05 | 0.87 ± 0.06 | 0.91 ± 0.06 | 0.90 ± 0.08 | 0.92 ± 0.07 | 0.8 ± 0.1 | 0.85 ± 0.07 |

Table 2: Results of multiple variations of the U-Net algorithm. The test cases differ by whether images were downsampled, or patches were extracted. The patch extraction cases differ by the patch size and whether overlapping patches were used for testing. The strides were the same for the horizontal and vertical directions. "None" indicates that no overlapping patches were used. AUC ROC is the area under the receiver operating characteristic curve, and AUC Precision-Recall is the area under the precision-recall curve.

## IV. Discussion

### IV.I. The Holes in the U-Net's Segmentation

The pathologist who contoured the digital pathology images in this study, identified the holes in the large tumor region in the U-Net's segmentation (Figures 3 and 5) as stroma, which is a non-tumoral supportive tissue, consisting mainly in fibroblasts, extracellular matrix, immune cells and vasculature. The algorithm can distinguish between epithelial cancer cells and these supportive tissues. The pathologist delineated as tumor areas, large clusters with a majority of tumor cells but often encased also non-tumor cells such as immune cells. Interestingly, the U-Net did not learn to make this approximation and created a more precise contour of the large tumor regions instead of excluding the stroma regions, hence creating the holes in the tumor regions. This demonstrates the ability of an algorithm that was trained on somewhat imprecise training data to learn a task and perform it with a higher level of precision than that of the training data from which it learned. While this was only observed for the segmentation of stroma, it might also be true for other features that are difficult to segment as well.

Although identifying the stroma regions is crucial as stroma tissue affect the growth and progression of tumors,[37] the algorithm received a negative score for delineating these regions. Specifically, the sensitivity of the U-Net is lower than it should be because of the stroma, which the U-Net classified as non-tumor, while they were classified as tumor in the ground truth data, thus resulting in a higher false negative rate. This might explain the relatively high variability in the sensitivity scores.

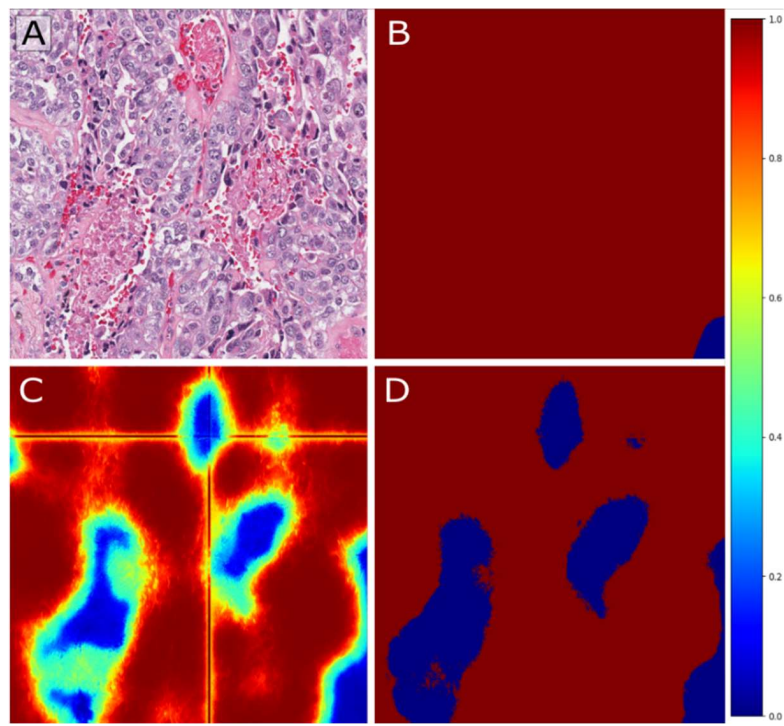

Figure 5: Zoomed in example of the stroma segmented in the image presented in Figure 3. A – The image used as input for the U-Net. B – The contour made by the pathologist. Tumor regions are contoured in dark red. All other regions are contoured in dark blue. Note, the pathologist contoured the stroma as part of a tumor region. C – Prediction map generated by the test case with 1872x1872 pixel patches and 939-pixel strides (case 2). D – The prediction map after applying a 50% confidence threshold.

### IV.II. Interpretation of Segmentation Scores

The similarity in the segmentation scores shows that the loss of information when downsampling the image affects the results as little as the contextual information withheld from the algorithm due to patch extraction. At an isotropic resolution of 1875x1875 pixels per image, tumor regions are still detectable.

However, the contours that were created using this method had a lower resolution than the original contours made by the pathologist.

Although one would expect the scores to get worse with smaller patches, the fact that there is no significant change in the scores when using small patches as opposed to large patches demonstrates that the U-Net models need very little contextual information to identify tumor regions. The patterns that make tumor distinguishable still seem clearly defined at a 464x464 pixel scale.

However, for larger tumors, edges are not as continuous when using smaller patches, as can be seen in Figure 4 for cases 5 and 6. The contours appear shifted along the edges of the square patches as opposed to the continuous borders observed in cases with larger patches (see Figure 4, cases 1 and 2). However, this does not affect any of the scores employed in this study. Small patches might not be well suited for further studies investigating edge detection.

The visually distinct edges of the patches in the reconstructed contours do not affect segmentation scores significantly and thus overlapping patches do not improve the scores significantly. However, the edges of the contours are more continuous when using overlapping patches. Using overlapping patches might therefore be recommended for studies involving edge detection.

### IV.III. Interpretation of the Segmentation Times

Generally, the trend in the segmentation time per image is that in test cases where algorithms compute less image space, they take less time. For example, the test case with downsampled images (case 7) only segments 1872x1872 pixels, whereas test case 6 segments 464x464x225 pixels.

### IV.VI. Comparison to Literature

The U-Net models explored in this study show similar results as deep convolutional neural networks that perform semantic tumor segmentation on histopathological core images from literature. For instance, Linmans et al. uses a modified U-Net architecture that exploits rotation and reflection symmetries and achieves a dice score of 83.7 when performing pixelwise segmentation on the PatchCamelyon dataset.[31] Further algorithms exist for tumor segmentation in histopathological images of lung tissue with adenocarcinoma such as the method developed by Wang *et al.*, which uses the modified fully convolutional network, ScanNet,[38] to perform patch-wise classification of whole slide images and achieves an accuracy of 0.820 and an AUC of 0.856 on the TCGA dataset.[30] Our method does not only achieve similar scores on a pixel-based level but also demonstrates the viability of using an "out of the box" architecture to contour tumor regions in histopathological images for future use in microdosimetry or theragnostic studies.

The advantage to using the well-known U-Net architecture is that it is not only readily available with numerous implementations available on platforms like GitHub but also that it is ready to use after a simple installation with little to no needed modifications for the specific usage case of histopathological image analysis. In addition to this, due to the U-Net's popularity there exist a multitude of helpful resources and documentation online for the usage of U-Net algorithms.

### IV.V. Limitations and Future Work

### IV.V.I. Inter-Observer Variability

While the proposed algorithms eliminate intra-observer variability of tumor segmentations, all training data was contoured by a single pathologist and are therefore prone to inter-observer variability. To create a less biased and more robust algorithm, contoured data from several pathologists could be used for training and testing.

Furthermore, an investigation quantifying the inter-observer variability between pathologists could be conducted as this would give insights into the natural limitations of tumor segmentation of histopathological images.

### IV.V.II. Stroma

Using data that only has epithelial tumor cells of adenocarcinoma delineated as tumor regions as opposed to including the stroma in the tumor regions, in cases where such distinctions are feasible, might not only improve segmentation scores such as the sensitivity but might also increase the accuracy and robustness of the U-Net models by teaching them to exclude stroma.

### IV.V.III. Limit Testing

A minimum patch size and a minimum image resolution should exist at which tumor cells can no longer be distinguished from non-tumor cells. While this study investigated patch sizes and image resolutions well above these limits, future studies might be done to investigate the performance of the U-Net near these limits. These studies might be motivated by trying to achieve the best image segmentation with the minimum amount of required computational power. For this study, 4 GB of memory is a low enough minimum requirement to make the proposed method widely applicable.

IV.V.IV. Clinical Use

The U-Net's accuracy and robustness as well as its low computational time make it suitable as a tool in a clinical setting. The proposed segmentation method could be used to assist pathologists when contouring histopathological images by serving as a "first draft" of the contours, which the pathologist can then tweak if they find any mistakes. The feasibility and efficiency of this workflow could be investigated in future studies.

## V. Conclusion

The well-known and readily available U-Net architecture can be used with the proposed pre-processing steps to accurately and consistently replicate tumor contours made by a pathologist up to $370 \pm 3$ times faster than manual contouring. This automation drastically reduces the time required to create the contours as well as eliminates access to pathologists and intra-observer variability in the segmentations. This method can be readily used for accurate and efficient tumor segmentation on histopathological core images for further studies in fields such radiotherapy dosimetry including microdosimetry and theranostics, thereby alleviating the bottleneck and dependency upon pathologist's time.

## VI. Acknowledgments

The authors would like to thank Ximeng Mao, Peter Kim and Majd Antaki at McGill University, Medical Physics unit for careful reading of the manuscript, their comments and insightful suggestions that increased the quality of our work.

This research was undertaken, in part, thanks to funding from the Canada Research Chairs Program (#253046 and 253047). Cette recherche a été menée dans le cadre des activités de l'Institut TransMedTech, grâce, en partie, au soutien financier des Fonds de recherche du Québec. This research was conducted as part of the TransMedTech Institute's activities and thanks, in part, to funding from the Canada First Research Excellence Fund. Computations were performed on the Niagara supercomputer at the SciNet HPC Consortium. SciNet is funded by: the Canada Foundation for Innovation; the Government of Ontario; Ontario Research Fund - Research Excellence; and the University of Toronto.

## VII. Conflicts of Interest

The authors have no conflict of interest to declare.

4 May 2021